\documentstyle[sprocl,rotating]{article}
\input{epsf}

\def\LC{LC}

\def\be{\begin{equation}}
\def\ee{\end{equation}}
\def\bea{\begin{eqnarray}}
\def\eea{\end{eqnarray}}
\newcommand{\beq}{\begin{equation}}
\newcommand{\eq}{\end{equation}}

\newcommand{\zb}{\bar{z}}

\newcommand{\intfeyn}{\int\limits_0^1}

\begin{document}

\title{HARD EXCLUSIVE QCD PROCESSES AT THE LINEAR COLLIDER
\footnote{Talk presented at LCWS 04}\\}

\author{ B.~PIRE$^A$,\,
 L.~SZYMANOWSKI$^B$ AND  S.~WALLON$^C$ }

\address{${}^A$\,CPHT \footnote{Unit{\'e} mixte 7644 du CNRS}, {\'E}cole
Polytechnique, 91128 Palaiseau, France \\[0.5\baselineskip]
${}^B$\,Soltan Institute , Warsaw, Poland \\[0.5\baselineskip]
${}^C$\,LPT \footnote{Unit{\'e} mixte 8627 du CNRS}, Orsay, France\\}

%%%%%%%%%%%%%%%%%%%%%%%%%%%%%%%%%%%%%%%%%%%%%%%%%%%%%%%%%%%%%%
% You may repeat \author \address as often as necessary      %
%%%%%%%%%%%%%%%%%%%%%%%%%%%%%%%%%%%%%%%%%%%%%%%%%%%%%%%%%%%%%%

\maketitle\abstracts{
The next generation of $e^+e^--$colliders will offer a possibility of
clean testing of QCD dynamics. Recent progress in the theoretical
description 
of exclusive processes permits
for many of them a consistent use of the perturbative QCD methods.
 We find that already on the basis of Born approximation,  the exclusive
diffractive production of two $\rho$ mesons
from virtual photons at very high energies should be measurable at the
linear collider (LC).
}

%***********************************************************************
\section{Introduction}
%***********************************************************************

The high energy limit of strong interaction has a very long story, which 
started much before the development of QCD \cite{revue}. The Regge {\it limit} corresponds to 
the kinematical regime of
large scattering energy square $s$ and small momentum transfer square $t$,
$s \gg -t$.
%It was first investigated in
%the
%context of Hadron-hadron experiments. 
The Regge {\it model} states
that the amplitude $A_{el}(s,t)$ of elastic hadron-hadron collision can be expressed as a sum over 
amplitudes corresponding
to the exchange of Regge trajectories in the $t$ channel.
 After partial wave expansion,
% their contribution to the amplitude is a pole.
they can be understood as states having continuous angular
momentum $\alpha_i(t)$ ($i$ labels the trajectory).
Such an hypothesis was supported experimentally by the famous
Chew-Frautschi diagram, where one could see,  when plotting spin
 as a function of mass square of known resonances,
an impressive alignement of linear trajectories. 
 It is a challenge for QCD to explain this experimental fact. 

Through the optical theorem, it is possible to relate the total
hadron-hadron cross-section with the imaginary part of the
forward elastic amplitude. Thus, $\sigma_{tot} \simeq
s^{\alpha_P(0)-1},$
where  $\alpha_P(t)$ is the Pomeron trajectory named after Pomeranchuk,
which is defined as the one carrying vacuum quantum
numbers. $\alpha_P(0)$
is called the intercept of the Pomeron trajectory.

Using unitarity and analyticity of $S$ matrix, Froissart could prove the
bound 
  $\sigma_{tot} \leq const \, ln^2 s \,.$

Since old studies of Donnachie and Lanshoff it is known that a
satisfactory description of the elastic and inelastic
hadronic data requires the soft Pomeron trajectory
$\alpha_P(t)=1 + 0.08 + 0.25 t,$ which explicitly violates Froissart
bound.
Thus, a need for unitarization was already present in the context of
Regge models.
 
Soon after QCD was proposed as a theory for strong interactions,
its Regge  limit was studied by Balitsky, Fadin, Kuraev and 
Lipatov \cite{bfkl}.
The evaluation of the elastic scattering amplitude of two 
infrared safe objects was performed, as an infinite series in $\alpha_s
\, ln s\,.$ This so-called Leading Log Approximation (LLA), where small values
 of perturbative $\alpha_s$  are compensated by large values of $\ln s,$
is expressed as an effective ladder with two reggeized gluons in
$t$-channel (gluons dressed by interaction, 
resulting in appearence of Regge trajectories) interacting with
$s-$channel
gluonic rungs,
through the effective Lipatov vertex which generalizes the usual
triple Yang-Mills vertex. The net result for this {\it hard} Pomeron intercept
is $\alpha_P(0)= 1+ c \, \alpha_s,$ where $c$ is a stricly positive constant,
which thus leads to a violation of the Froissart bound at perturbative
level.
Such an approach has intrinsic limitations. $\alpha_s$ is fixed in
the LLA approximation. Along the effective ladder, typical transverse
momenta of reggeons diffuse into the IR domain, with a typical gaussian
shape, the so-called Bartels cigar, which broadness
increases
with $s.$ Higher order correction in the N(ext)LLA approximation have
been computed. They are large and highly dependent on the choice of 
scale of the running coupling constant. Thus, various resummation schemes have
been
proposed, in order to compute the effective Pomeron intercept from QCD.
Because of the explicit violation of the Froissart bound by the BFKL
Pomeron,
an intense activity is now devoted to the problem of unitarization of
QCD, and to the related problem of saturation, 
which avoids the unlimited growth of gluon density
with increase of $s$.

\section{Phenomenology of QCD Pomeron}

In order to test the hard Pomeron, it is not enough to study large $s$
experiments. It is also compulsory to select processes where a hard
scale enables one to use of perturbative QCD. Such an applicability is 
more intricate than for conventional QCD evolution \cite{dglap}. 
Indeed, the usual Operator Product Expansion is not
any more valid in the Regge limit of QCD, and one needs to use some
generalized version of QCD factorization. In order to emphazise the
effects
of infrared singularities of QCD (responsible for soft
Bremstrahlung effects) with respect to collinear singularities (which are
the
source of the conventional DGLAP evolution), processes with comparable
characteristic scales at both end
of the effective Pomeron ladder have been objects of special interests.
In hadron-hadron colliders (Tevatron , LHC), processes with inclusive 
production of two high $p_t$ jets with large relative rapidity $Y$
(related to $s$ by $Y=\ln s/s_0$), known as Mueller-Navelet jets, give
access
to the hard Pomeron at $t=0.$
Diffractive high energy jet production, with a large gap in rapidity
between the two jets (no activity in the detector between them), at
large $t$ (which provides the hard scale), reveals the Pomeron
structure
at large $t.$

In DIS, the virtuality of the photon naturaly provides a hard scale.
At the level of both total and diffractive cross-sections,
 it was possible to describe HERA data using models based on
BFKL
type of evolution, although the distinction with  standard DGLAP
evolution is not conclusive \cite{F2}.
Exclusive vector meson production was also proposed in
order
to see BFKL effects, selecting events with a large gap in rapidity
between
the vector meson and the proton remnants.
These approaches needed however some ansatz for the coupling of the
proton-Pomeron
coupling.
%Energetic forward jet production in DIS naturally provides a second
%scale, and can reveal the $t=0$ BFKL dynamics. Such an
%effect 
%however depends on the choice of scale of the coupling.

\section{$\gamma^*\gamma^*$ processes: the gold plated experiment}

Each of the phenomenological tests described above 
have various limitations, mainly
related to the fact that non-perturbative inputs are always needed.

From the theoretical point of view, 
the best way for studying typical Regge behaviour in perturbative
QCD is provided by the scattering of small transverse size objects. 
Such a reaction is naturally provided by a photons of high
virtuality as produced in $e^+e^-$ tagged collisions.
This was investigated at the level of total  $\gamma^*
\gamma^*$
cross section  
by various groups \cite{bfklinc}. Typical Pomeron
enhancement can hardly be seen at LEP, but should be definitely measurable at
\LC.
One of the key point in order to reveal this effect is that the detectors
should be able to tag the outgoing particle with minimal tagging angle down to
20 mrad.

Another possibility is to select specific $heavy$ bounds states
($J/\Psi,\Upsilon,...$) in the final
state. 
This has been studied in the case of double diffractive
photo production of $J/\Psi$ \cite{jpsi}.
Several tens of thousand events are expected at \LC,
with an enhancement factor of the order of 50 with respect to the Born 
estimate.

We study the process
of exclusive electroproduction
of two  $\rho-$mesons in the $\gamma^* \gamma^*$
collisions. 
The virtualities  $Q_1^2$ and $Q_2^2$ of the scattered photons play
 the role of the hard 
scales. 
 This allows one to  scan $Q_1^2,$ $Q_2^2,$
as well as  $t$ to test the structure of the hard Pomeron. It is also possible to study various polarizations of both photons and
mesons.
As a first step in this direction we shall consider this
process with  longitudinally polarized photons and 
$\rho-$mesons,
 \beq
\label{process}
\gamma_L^*(q_1)\;\gamma_L^*(q_2) \to \rho_L(k_1)  \;\rho_L(k_2)\;.
\ee
 The choice of longitudinal polarizations of both the
scattered photons and produced vector mesons is dictated by the
fact that this configuration of the lowest twist-2 gives the dominant
contribution in the powers of the 
hard scales $Q_{1,2}^2$. As a guiding line, one
should remember the HERA data where the cross-section of diffractive  photoproduction of $J/\Psi$
is comparable to the  cross-section of diffractive  electroproduction of
$\rho$ when the virtuality of the photon is of the order of the mass squared
of the $J/\Psi$, which can easily be understood heuristically by
crossing symetry arguments. So one may guess that
 $\gamma^*\gamma^* \to \rho\rho$
and $\gamma\gamma \to J/\Psi \,J/\Psi$ cross-sections to be comparable
at $Q_1^2=Q_2^2 \sim m_{J/\Psi}^2.$

\begin{figure}[htp]
\epsfxsize=5.0cm{\centerline{\epsfbox{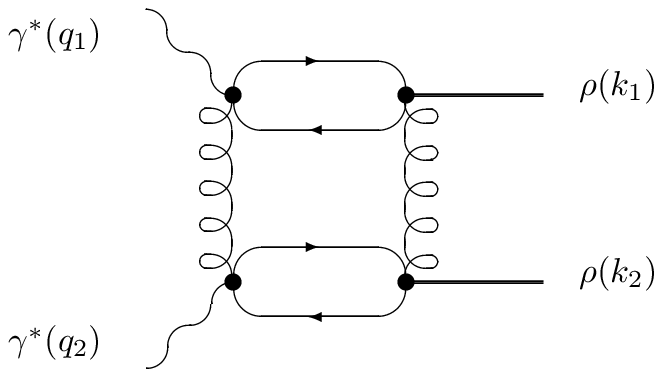}}}
\caption{Amplitude for the process $\gamma^*\gamma^* \to \rho\rho$
at Born order. The dots denote the effective coupling of t-channel gluons to
the impact factors. Virtualities are defined by $Q_{1(2)}^2=-q_{1(2)}^2.$}
\label{dessinprocess}
\end{figure}

We compute the Born order contribution to the process (\ref{process}) using
the impact representation, as illustrated in Fig.\ref{dessinprocess}.

The meson vertex is treated in the collinear approximation which
neglects in the hard part of the amplitude the relative transverse
momentum of the quarks. This results in appearence of the
Distribution Amplitude (DA): the
meson wave function integrated over the relative momentum of quarks.

The amplitude for the process reads
\beq
{\cal M}= -i\,  4 \pi \, \alpha_{em} \,  s \, \alpha_s^2 \,f_{\rho}^2\, Q_1\, Q_2
\frac{C_F}{N_c} \intfeyn \intfeyn  dz_1 \, dz_2 \, z_1 \, \zb_1 \, z_2 \, \zb_2
\, \Phi(z_1) \, \Phi(z_2) M(z_1,z_2)\,
\eq
where $\Phi(z)=6 z \zb$ is the asymptotic DA of the $\rho$ meson, $z \,(\zb)$
being the light-cone fraction of the $\rho$ momentum 
carried by the quark (resp. antiquark).  
$M(z_1,z_2)$ is the transverse momentum convolution of the impact factors with 2 $t-$channel gluon propagators.
It can be expressed through two integrals with respectively 3 propagators (1 massive, 2 massless) and 4 propagators (2 massive with different masses, 2 massless), These two integrals were computed exactly using a generalized version of a technique used in coordinate space when evaluating 
diagrams of massless two dimensional conformal field theories. 

\begin{figure}
\epsfxsize=5.0cm{\centerline{\epsfbox{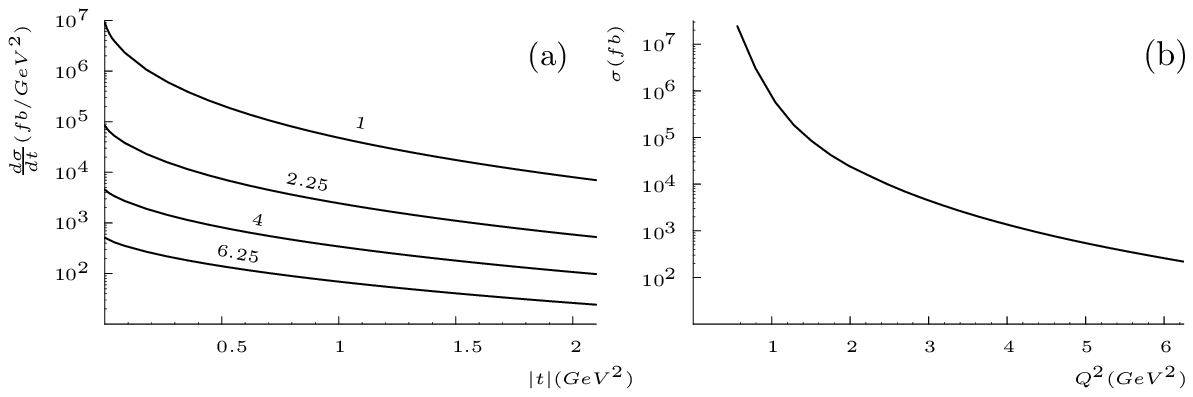}}}
\caption{Born order result for (a)\,$\frac{d \sigma^{\gamma^*\gamma^*\to \rho
  \rho}}{dt}$ as a function of  $|t|\,(GeV^2)$ for various
  values of $Q^2\,(GeV^2)$, (b)\, $\sigma^{\gamma^*\gamma^*\to \rho \rho}$ as a function of $Q^2\,(GeV^2)$. }
\label{resultat}
\end{figure}

The result for $M(z_1,z_2)$ is regular in $z_1$ and $z_2.$ After numerical integration over $z_1$ and $z_2$ and squaring, one obtains the 
differential cross-section 
$\frac{d \sigma^{\gamma^*\gamma^*\to \rho \rho}}{dt},$ shown 
%\beq
%\frac{d \sigma^{\gamma^*\gamma^*\to \rho \rho}}{dt}=\frac{|{\cal M}|^2}{16 \pi s^2}\,.
%\eq 
in Fig.\ref{resultat}a,
 for various values of $Q_1^2=Q_2^2=Q^2.$
It is rapidly decreasing in $t$, and flat in $s.$ Any BFKL type of resummation
would give a rising shape in $s.$ 
Integrating over $t,$ one gets the $\sigma^{\gamma^*\gamma^*\to \rho \rho}$
cross-section.
 As can be seen from Fig.\ref{resultat}b,
it is a power like decreasing function of $Q,$ as $1/Q^{10.5}.$
This is due to the impact factors structure. 
%Note that is compatible with HERA data, where the shape of the cross-section
% for diffractive electroproduction of $\rho$
%looks like $1/Q^{4.5}.$ 

The expected number of events at \LC, for a nominal luminosity of $100 fb^{-1},$ is of the order of 1000 events per year.
This is only a lower bound since the contribution of the transverse
photon case is to be added.
Morover, we expect  a net and visible enhancement of this cross section,
because of resummation effects \`a la BFKL. 

\section{Conclusions}

Double diffractive $\rho$ production in $e^+e^-$ collisions is 
a crucial test for QCD in Regge limit.
The Born contribution for longitudinally polarized photon and meson
 gives a measurable cross-section.
BFKL enhancement remains to be evaluated.

$e^+e^-$ collisions would be also a very good place to observe and test the Odderon. Such an object is the partner of the Pomeron, 
with opposite charge conjugation. We propose to study double diffractive $\pi^0$ production from two highly virtual photons,
which should be dominated by the $t-$channel exchange of an Odderon. In
QCD, such a state is constructed from at least 3 gluons, and resummation
effects are expected in the Regge limit \cite{ewerz}.
To test the existence of Odderon at the amplitude level,
one may study interference effects between Odderon and Pomeron exchange
in  $\gamma^*\gamma^*\to (\pi^+ \pi^-) (\pi^+ \pi^-)$ processes, using the
fact that the C-parity is not fixed for such final 
states \cite{pire}. 

\section*{Acknowledgment} The work of L.Sz. is partially supported by
the French-Polish scientific agreement Polonium.
%1)
%     \begin{figure}[ht] 
%    \begin{center}
%     \vspace*{.2cm}
%     \epsfig{file=j_aleph_KK.eps,width=6.5cm}
%     \end{center}
%     \caption{$\tau \rightarrow \nu_\tau K_s K_l ~\pi$.} 
%     \label{fig:tauKK}
%     \end{figure}
%2)
%     \begin{figure}[ht]
%     \begin{center}
%     \begin{tabular}{cc}
%     \mbox{\epsfig{file=wwzz30_colourH.eps,width=5.7cm}}&
%     \mbox{\epsfig{file=wwzz60_colourH.eps ,width=5.7cm}}
%     \end{tabular}
%     \end{center}
%     \caption{Study for the $W_L$ scattering. Pairs of W's and Z's have been
%generated with different jet resolutions. The resolution is parametrised as
%$\Delta E = \alpha / \sqrt E$, on the left $\alpha = 0.3$,
%on the right $\alpha = 0.6$.}
%     \label{fig:wwzz}
%     \end{figure}

%References are denoted and numbered as [1], [2], or [3-5] at the end of 
%the referring words or sentences [6].  

%\section{Conclusion}

%..... but instead of what?

\end{document}